\documentclass[prd,twocolumn,preprintnumbers,amsmath,amssymb,nofootinbib]{revtex4}
\newcommand{\bg}{\bar{g}}
\newcommand{\Eqref}[1]{Eq.~\eqref{#1}}
\usepackage{color}

\usepackage{amsmath}
\usepackage{bbm}
\usepackage{amssymb}
\usepackage{slashed}
\usepackage{graphicx}
\usepackage{graphics}
\usepackage{epsfig}
\usepackage{hyperref}

\usepackage{color}
\definecolor{comment}{rgb}{0.9,0,0}
\definecolor{holger}{rgb}{0,0.5,0.7}

\begin{document}

\title{Ghost anomalous dimension in asymptotically safe quantum gravity}
\date{\today} \author{Astrid Eichhorn and Holger Gies}

\affiliation{\mbox{\it Theoretisch-Physikalisches Institut, Friedrich-Schiller-Universit{\"a}t Jena,}
\mbox{\it Max-Wien-Platz 1, D-07743 Jena, Germany}
\mbox{\it E-mail: {astrid.eichhorn@uni-jena.de, holger.gies@uni-jena.de}}
}

\begin{abstract}

We compute the ghost anomalous dimension within the asymptotic-safety scenario for quantum gravity. For a class of covariant gauge fixings and using a functional RG scheme, the anomalous dimension $\eta_c$ is negative, implying an improved UV behavior of ghost fluctuations. At the non-Gau\ss ian UV fixed point, we observe a maximum value of $\eta_c\simeq -0.78$ for the Landau-deWitt gauge within the given scheme and truncation. Most importantly, the backreaction of the ghost flow onto the Einstein-Hilbert sector preserves the non-Gau\ss ian fixed point with only mild modifications of the fixed-point values for the gravitational coupling and cosmological constant and the associated critical exponents; also their gauge dependence is slightly reduced. Our results provide further evidence for the asymptotic-safety scenario of quantum gravity.

\end{abstract}
\maketitle
\section{Introduction}
%
The asymptotic-safety scenario, introduced by Weinberg in 1979
\cite{Weinberg:1980gg,Weinberg:1996kw,Weinberg:2009ca}, provides for a
promising route to a fully consistent and predictive theory of quantum
gravity. Purely within the framework of standard local quantum field theory,
gravity can become renormalizable at an interacting non-Gau\ss ian fixed point
in the ultraviolet (UV). A verification of this scenario requires a
non-perturbative analysis of the renormalization flow induced by the
fluctuations of the gravitational degrees of freedom of the theory.

Convincing evidence for this scenario has been collected in the past years,
using the metric field itself as the microscopic degree of freedom. Based on
the pioneering work of Reuter \cite{Reuter:1996cp}, the renormalization flow
of generic gravitational theories including Einstein-Hilbert terms and
higher-order curvature invariants has been studied within the functional RG
\cite{Dou:1997fg,Reuter:2001ag,Lauscher:2002sq,Lauscher:2001cq,Lauscher:2001ya,Lauscher:2001rz,Souma:1999at,Reuter:2002kd,Bonanno:2004sy,Reuter:2007rv,Lauscher:2005xz,Percacci:2002ie,Codello:2006in,Codello:2007bd,Codello:2008vh,Litim:2003vp,Fischer:2006fz,Machado:2007ea,Reuter:2008wj,Reuter:2005bb,Percacci:2007sz,Benedetti:2009rx, Daum:2008gr} and other field-theoretical methods
\cite{Niedermaier:2003fz,Niedermaier:2009zz}. These results do not only agree
on the existence of a non-Gau\ss ian fixed point, but also on the classification of renormalization
group (RG) relevant and irrelevant directions. In particular, the number of RG
relevant directions corresponds to the number of physical parameters to be
fixed for the theory to be fully predictive; there are strong indications from
high-order calculations that this number is finite
\cite{Codello:2007bd,Machado:2007ea,Benedetti:2009rx}, lending substantial support to the
scenario that Quantum Einstein Gravity (QEG) is asymptotically safe and
applicable to diverse physical scenarios. For instance, the implications of
QEG for cosmology as well as black hole physics have been studied in
\cite{Bonanno:2008xp,Weinberg:2009wa,Reuter:2005kb}. Some features of the
asymptotic-safety scenario are also reminiscent to results obtained within other
nonperturbative approaches to a quantum field theory of gravity
\cite{Ambjorn:2001cv, Ambjorn:2005qt, Ambjorn:2009ts, Smolin:1981rm}.

The concept of asymptotic safety, of course, extends to general quantum field
theories: for instance, corresponding scenarios curing the triviality problem
of the Higgs sector of the particle-physics standard model have been developed
in the pure matter sector \cite{Gies:2009sv,Gies:2009hq} and also including
gravity \cite{Zanusso:2009bs,Shaposhnikov:2009pv}.

As for any quantum field theory with a local symmetry group, a continuum
formulation based on the gauge-variant degrees of freedom requires gauge
fixing conventionally accompanied by Faddeev-Popov ghosts. Whereas most
studies have concentrated on gauge-invariant building blocks of the theory,
the investigation of the renormalization flow of the ghost sector has started
only recently \cite{Eichhorn:2009ah}. Even though the ghosts do not belong to
the physical state space, their fluctuations can, in principle, contribute in
an important manner to gauge-invariant observables, depending on the details
of the gauge. In fact, in the Landau gauge in Yang-Mills theory, the ghost
sector can even carry crucial physical information in the strongly interacting
regime of the theory, as is, for instance, expressed in the Kugo-Ojima or
Gribov-Zwanziger scenarios for confinement: here, signatures for the physical
property of confined color charges occur in the form of a strongly divergent
ghost propagator in the infrared, whereas the gluon propagator is finite or
even zero
\cite{Kugo:1979gm,Gribov:1977wm,Zwanziger:2003cf,Alkofer:2000wg,Fischer:2006ub}. These
properties can be related to the existence of Gribov copies, i.e., the
non-uniqueness of the gauge-fixing condition and the related zeros of the
Faddeev-Popov operator \cite{Gribov:1977wm, Zwanziger:2003cf}. This problem
also exists in gravity \cite{Das:1978gk,Esposito:2004zn}, where it is even
more challenging due to the highly non-trivial topology of the diffeomorphism
group.

As an indicator for the relevance of ghost fluctuations at a UV fixed point of
asymptotically safe Quantum Gravity, we compute the ghost anomalous dimension in this work. On the one hand
this measures how these gauge-variant degrees of freedom
propagate in a given gauge, on the other hand the ghost backreaction on the
 renormalization flow of the graviton sector provides for another nontrivial test of the
existence of the UV fixed point.

Such investigations require a non-perturbative tool which supports an approach
to quantum gravity within systematic and consistent approximation
schemes. Such an advanced framework is provided by the functional
renormalization group which facilitates studies not only of a given action but
of the renormalization flow of generic effective actions in theory space (for
reviews for gauge theories, see \cite{Reuter:1996ub}). The
functional RG can be formulated in terms of a flow equation for a
scale-dependent effective average action $\Gamma_k$  with $k$ denoting a momentum
scale. The Wetterich flow equation \cite{Wetterich:1992yh}
relates the flow of $\Gamma_k$ to the trace of the full non-perturbative
propagator,
\begin{equation}
 \partial_t \Gamma_k = \frac{1}{2} {\rm STr} \frac{1}{\Gamma_k^{(2)}+R_k}\partial_t R_k,\label{floweqn}
\end{equation}
where $t= \ln k$ and $\Gamma_k^{(2)}$ denotes the second functional derivative
with respect to the fields. The regulator function $R_k$ acts as a momentum-
and $k$-dependent mass term and cuts off momentum modes with $p^2 \lesssim
k^2$. Accordingly, the limit $k \rightarrow \infty$ is related to the classical
action $S_{\rm cl}$ \cite{Manrique:2008zw}, whereas the full quantum effective
action $\Gamma$ is obtained for $k \rightarrow 0$.

As $\Gamma_k$ is a functional in the infinite dimensional space of all
operators compatible with the symmetry constraints, an exact solution
of the Wetterich equation is difficult. Approximate solutions can be found by
truncating a systematic expansion of the effective action. The reliability of
an approximation can be studied not only by checking its internal consistency
but also by actively monitoring the convergence of physical quantities at
increasing order in this expansion.

In order to extract gauge-invariant parts of the action functional, the
background-field formalism \cite{Abbott:1980hw} is most convenient. Here the
full metric $\gamma_{\mu \nu}$ is split into a background $\bar{g}_{\mu \nu}$
and a fluctuation metric $h_{\mu \nu}$:
\begin{equation}
 \gamma_{\mu \nu}= \bar{g}_{\mu \nu}+ h_{\mu \nu}.
\end{equation}
Accordingly, we denote covariant derivatives compatible with the full
or background metric by $D_{\mu}$ or $\bar{D}_{\mu}$,
respectively. In the present work, we study the following truncation of the
effective action in four dimensions:
\begin{equation}
 \Gamma_k = \Gamma_{k \,\rm EH}+ \Gamma_{k\,\rm gf}+ \Gamma_{k\, \rm gh},
\end{equation}
where 
\begin{eqnarray}
\Gamma_{k\,\mathrm{EH}}&=& 2 \bar{\kappa}^2 Z_{\text{N}} (k)\int 
d^4 x \sqrt{\gamma}(-R+ 2 \bar{\lambda}(k))\label{eq:GEH},\\
\Gamma_{k\,\mathrm{gf}}&=& \frac{Z_{\text{N}}(k)}{2\alpha}\int d^4 x
\sqrt{\bar g}\, \bar{g}^{\mu \nu}F_{\mu}[\bar{g}, h]F_{\nu}[\bar{g},h]\label{eq:Ggf},
\end{eqnarray}
with
\begin{equation}
 F_{\mu}[\bar{g}, h]= \sqrt{2} \bar{\kappa} \left(\bar{D}^{\nu}h_{\mu
   \nu}-\frac{1+\rho}{4}\bar{D}_{\mu}h^{\nu}{}_{\nu} \right). 
\end{equation}
Herein, $\bar{\kappa}= (32 \pi G_{\text{N}})^{-\frac{1}{2}}$ is related to the
bare Newton constant $G_{\text{N}}$. The ghost term with a
wave function renormalization $Z_c$ is given by
\begin{eqnarray}
 \Gamma_{k\, \rm gh}&=& -\sqrt{2} \int d^4x \sqrt{\bg}\, Z_c \,\bar{c}_{\mu} 
\Bigl(\bar{D}^{\rho}\bar{g}^{\mu \kappa}\gamma_{\kappa \nu}D_{\rho}\nonumber\\
&+& \bar{D}^{\rho}\bar{g}^{\mu \kappa}\gamma_{\rho \nu}D_{\kappa}
- \frac{1}{2}(1+\rho)\bar{D}^{\mu}\bar{g}^{\rho \sigma}\gamma_{\rho \nu}D_{\sigma} \Bigr)c^{\nu}.
\end{eqnarray}
In this work we neglect the possibility of a different flow for the two tensor
structures in the kinetic ghost term and focus on the single wave
function renormalization $Z_c$.

During the course of the flow, we also identify the full metric
$\gamma_{\mu\nu}$ with the background metric $\bar{g}_{\mu\nu}$, once the
propagator of the fluctuation field is determined. This approximation appears
justified for many aspects of the Einstein-Hilbert sector
\cite{Manrique:2009uh}. More generally, we expect bimetric truncations to be mandatory for
a computation of (momentum-dependent) propagators and vertices. For instance
in Yang-Mills theories, the distinction between fluctuation and background is
crucial for the computation of gauge-invariant infrared observables
\cite{Braun:2007bx}.

The paper is organized as follows: Sect.~\ref{sec:etac} summarizes our
  method to extract the ghost anomalous dimension and presents general results
  for a class of covariant gauges and a spectrally adjusted regulator. In
  Sect.~\ref{sec:results}, explicit results for the Landau-deWitt gauge and
  the deDonder gauge are given. We discuss the implications of our results and
  conclude in Sect.~\ref{sec:conc}. Many technical details and explicit
  representations may be found in the appendices.

\section{Ghost anomalous dimension}
\label{sec:etac}

In order to extract the anomalous dimension of the ghost, we project the flow
equation onto the running of the ghost wave function renormalization. We use a
decomposition of $\Gamma^{(2)}_k+R_k=\mathcal{P}+\mathcal{F}$ into an inverse
propagator matrix $\mathcal{P}=\Gamma_k^{(2)}[\bar{c}=0=c]+R_k$, including the
regulator but no external ghost fields, and a fluctuation matrix
$\mathcal{F}=\Gamma_k^{(2)}[\bar{c},c]-\Gamma_k^{(2)}[\bar{c}=0=c]$
containing external ghost fields. The components of $\mathcal{F}$ are either
linear or bilinear in the ghost fields, as higher orders do not occur in our
truncation. We may now expand the right-hand side of the flow equation as
follows:
\begin{eqnarray}
 \partial_t \Gamma_k&=& \frac{1}{2}{\rm STr} \{
 [\Gamma_k^{(2)}+R_k]^{-1}(\partial_t R_k)\}\label{eq:flowexp}\\
&=& \frac{1}{2} {\rm STr} \tilde{\partial}_t\ln \mathcal{P}
+\frac{1}{2}\sum_{n=1}^{\infty}\frac{(-1)^{n-1}}{n} {\rm
  STr} \tilde{\partial}_t(\mathcal{P}^{-1}\mathcal{F})^n,\nonumber
\end{eqnarray}
where the derivative $\tilde{\partial}_t$ in the second line by definition
acts only on the $k$ dependence of the regulator, $\tilde{\partial}_t=
\int \partial_t R_k\frac{\delta}{\delta R_k}$. Since each factor of $\mathcal {F}$ contains a
coupling to external fields, this expansion simply corresponds to an expansion
in the number of vertices.

To bilinear order in the external ghost and antighost, we may directly neglect
all contributions beyond $(\mathcal{P}^{-1}\mathcal{F})^2$. Diagrammatically, the remaining terms correspond to a tadpole and a self-energy diagram.
\begin{figure}[!here]
\begin{center} 
\includegraphics[scale=0.4]{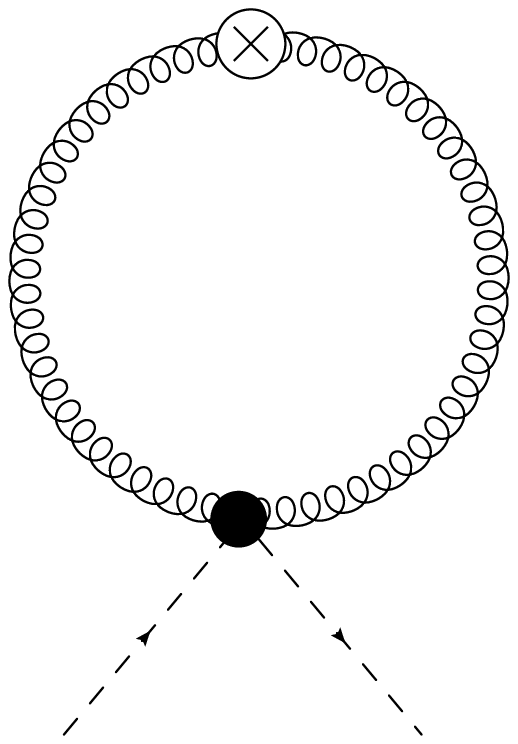}
\end{center}
\begin{minipage}{0.45\linewidth}
 \includegraphics[scale=0.45]{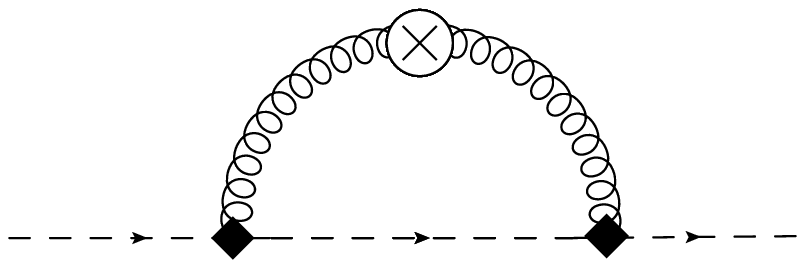}
\end{minipage}
\begin{minipage}{0.45\linewidth}
 \includegraphics[scale=0.45]{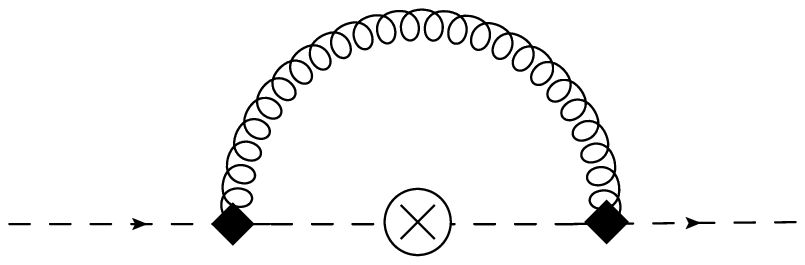}
\end{minipage}
\caption{Tadpole diagram with the two-graviton-ghost-antighost vertex denoted
  by a dot and self-energy diagrams with graviton-ghost-antighost vertex denoted by
  a diamond. The crossed circle represents the regulator insertion $\partial_t
  R_k$ which occurs on both of the two propagators for the self-energy
  diagram.}
\end{figure}
Irrespective of our choice of gauge in the ghost sector, i.e., for all values
of the gauge parameter $\rho$, we find that the tadpole does not contribute to
the ghost anomalous dimension as the corresponding vertex is zero; this is
shown in appendix \ref{tadpole}.

We take advantage of the possibility to specify a concrete background for
computational purposes in order to project unambiguously onto the running
couplings. For the running of the ghost wave function renormalization, it
suffices to choose a flat background, i.e. $\bar{g}_{\mu \nu}= \delta_{\mu
  \nu}$, as this allows for a clean separation between the kinetic ghost
operator and higher-order operators coupling ghost bilinears to curvature
invariants. It is important to stress that this choice of background has
nothing to do with a perturbative reasoning, where one would consider small
metric fluctuations around a flat background. For the projection of the
Wetterich equation, any convenient background that helps distinguishing
different operators unambiguously can be used in order to extract the
corresponding $\beta$ functions. In fact, we are allowed to choose different
backgrounds for extracting the flow of different operators. The notion of a
background in the sense of the true quantum expectation value of the field can
finally be computed, once the full effective action is known.

On a flat background, we project the flow equation in Fourier space onto
the running wave function renormalization by
\begin{eqnarray}
 \eta_c&=& -\partial_t {\rm ln}Z_c\label{projection} \\
&=&-\frac{1}{\sqrt{2}Z_c} 
\frac{1}{4}\delta^{\alpha \gamma}\frac{\partial}{\partial \tilde{p}^2}
\int \frac{d^4 \tilde{q}}{(2\pi)^4}
\left(\frac{\overset{\rightarrow}{\delta}}{\delta \bar{c}^{\alpha}(\tilde{p})} 
\partial_t \Gamma_k
\frac{\overset{\leftarrow}{\delta}}{\delta c^{\gamma}(\tilde{q})}\right).\nonumber
\end{eqnarray}
Our conventions for the functional Grassmannian derivatives are such that
\begin{equation}
 \frac{\overset{\rightarrow}{\delta}}{\delta \bar{c}^{\alpha}(\tilde{p})}
 \int \frac{d^4p}{(2 \pi)^4}\bar{c}^{\mu}(p)M_{\mu \nu}(p)c^{\nu}(p) 
\frac{\overset{\leftarrow}{\delta}}{\delta c^{\gamma}(\tilde{q})}
= \delta(\tilde{p},\tilde{q})M_{\alpha
  \gamma}(\tilde{p}), 
\end{equation}
where $\delta(p,q)=(2\pi)^4 \delta^{(4)}(p-q)$.  In the following, we
decompose the fluctuation metric into irreducible representations of the
Poincar\'{e} group in momentum space, i.e., transverse traceless ($h^{\text{T}}_{\mu \nu}$), transverse
vector ($v_{\mu}$), and scalar ($\sigma$ and $h$) degrees of freedom:
\begin{eqnarray}
 h_{\mu \nu}(p)&=& h_{\mu \nu}^{\text{T}}(p)+ i \frac{p_{\mu}}{\sqrt{p^2}}v_{\nu}(p)
+ i \frac{p_{\nu}}{\sqrt{p^2}}v_{\mu}(p)\nonumber\\
&+& \frac{p_{\mu}p_{\nu}}{p^2}\sigma(p)
-\frac{1}{4}\delta_{\mu \nu}\sigma(p)+\frac{1}{4}\delta_{\mu \nu}h(p) \label{York},
\end{eqnarray}
where $h(p)= \gamma^{\mu \nu}h_{\mu \nu}$ is the conformal mode.
In this decomposition of the graviton, the quantities $\mathcal{P}^{-1}$ and
$\mathcal{F}$ are $6\times6$ matrices in field space $\phi=(h^{\text{T}}_{\mu\nu},v_\mu,\sigma,h,\bar{c}_\mu,c_\mu)$. In the following,
  we confine ourselves to the gauge choice $\rho\rightarrow \alpha$ for which
  the propagator matrix is diagonal in the graviton modes and off-diagonal in
  the ghosts. Accordingly, the product
$\left(\mathcal{P}^{-1}\mathcal{F}\right)^2$ decomposes into four terms from
the four different graviton modes, which may be considered separately. The
corresponding matrix elements of $\mathcal{P}^{-1}$ and $\mathcal{F}$ which
build up the self-energy diagram can be decomposed into elementary propagators
and vertices for these modes, listed in detail in appendix
\ref{vertices}. Incidentally, the gauge choice $\rho\to\alpha$ becomes
problematic for $\rho=\alpha=3$: here, both the inverse ghost as well as the
inverse scalar graviton propagators develop a zero mode. The inverse
propagator is thus not invertible, signaling that the gauge fixing is no
longer complete for this choice. Nevertheless, the most important standard
gauge choices $\alpha=0$ or $\alpha=1$ are unaffected by this problem.

The matrix multiplication in \Eqref{eq:flowexp} to second order in the
vertices can be performed straightforwardly. Also, the trace over Lorentz
indices and momentum variables can efficiently be performed by standard means
and simplifies by using the projection rule as specified by
\Eqref{projection}. Finally evaluating the $\tilde{\partial}_t$ derivative
necessitates a more explicit form of the regulator. We choose a spectrally and
RG-adjusted regulator \cite{Gies:2002af,Litim:2002xm},
\begin{equation}
 R_k(p^2)= \Gamma_k^{(2)}(p^2) \, r \left(\frac{\Gamma_k^{(2)}(p^2)}{Z k^2} \right).
\end{equation}
Herein, the wave function renormalization $Z$ in the regulator
is adapted to each mode, such that the cutoff scales for each mode are equal;
this implies that the factor of $Z$ also contains the numerical prefactors
from \Eqref{Gamma2EHYork} for the transverse, vector, scalar and conformal
mode.  This adjustment is useful, as different effective cutoff scales for
different modes could artificially alter the results within a truncation. 

We thereby arrive at the following flow equation, where the first line is the
transverse traceless contribution. The second line is due to the transverse
vector mode, and the last two lines result from the two scalar modes,
respectively. The terms $ \sim r'$ are due to the external momentum $\tilde{p}$ flowing through the internal ghost line and being acted upon by the $\partial_{\tilde{p}^2}$-derivative in \Eqref{projection}. 
\begin{widetext}
\begin{eqnarray}
\eta_c &=&\sqrt{2} Z_c\tilde{\partial}_{t}  \int \frac{dp^2}{16 \pi^2}p^2\,  \Biggl[ \frac{5(\alpha-7)}{18 (\alpha-3) \bar{\kappa}^2Z_{\text{N}} \left(p^2-2\bar{\lambda}(k)\right)\left(1+r\left(\frac{p^2-2 \bar{\lambda}(k)}{k^2} \right) \right) \sqrt{2}Z_c \left(1+r\left(\frac{p^2}{k^2} \right) \right)}\nonumber\\
&{}&+\frac{ \alpha }{\bar{\kappa}^2Z_{\text{N}}\left(p^2-2\bar{\lambda}(k)\right)\left(1+r\left(\frac{p^2-2 \bar{\lambda}(k)}{k^2} \right) \right)\sqrt{2}Z_c \left(1+r\left(\frac{p^2}{k^2} \right) \right)}\left(\frac{1}{3}-\frac{1}{4}\frac{p^2}{k^2}\frac{r'\left(\frac{p^2}{k^2}\right)}{1+r\left(\frac{p^2}{k^2} \right)} \right)
\label{eq:13}\\
&{}&-  \frac{\alpha}{18 (\alpha-3)\bar{\kappa}^2Z_{\text{N}}\left((\alpha-3)p^2+4 \alpha\bar{\lambda}(k)\right)\left(1+r\left(\frac{p^2+\frac{4 \alpha \bar{\lambda}(k)}{\alpha-3}}{k^2} \right) \right)\sqrt{2}Z_c \left(1+r\left(\frac{p^2}{k^2} \right) \right)}\left(2 (\alpha-7)+3 \frac{p^2}{k^2}\frac{r'\left(\frac{p^2}{k^2}\right)}{1+r\left(\frac{p^2}{k^2} \right)}\right)\nonumber\\
&{}&+ \frac{ 1}{6 (\alpha-3)\bar{\kappa}^2Z_{\text{N}}\left((\alpha-3)p^2+4 \bar{\lambda}(k)\right)\left(1+r\left(\frac{p^2+\frac{4 \bar{\lambda}(k)}{\alpha-3}}{k^2} \right) \right)\sqrt{2}Z_c \left(1+r\left(\frac{p^2}{k^2} \right) \right)}\left((3 \alpha-4)-\alpha\frac{p^2}{k^2}\frac{r'\left(\frac{p^2}{k^2}\right)}{1+r\left(\frac{p^2}{k^2} \right)} \right)
\Biggr]\nonumber
\end{eqnarray}
\end{widetext}
The Landau-deWitt gauge $\alpha =0$ clearly plays a distinguished role
as only the transverse traceless and the conformal mode propagate. This
feature appears generically in all diagrams with more than one vertex as the
propagator is $\sim \alpha$. This favors the Landau-deWitt gauge from a
computational point of view. As it is moreover a fixed point of the
renormalization group flow \cite{Ellwanger:1995qf,Litim:1998qi}, we consider
the Landau-deWitt gauge as our preferred gauge choice.

\section{Results}
\label{sec:results}
For explicit computations, we use an exponential regulator shape function
$r(y)$, 
\begin{equation}
 r(y)=\frac{1}{e^{y}-1}.
\end{equation}
Such a smooth shape function is advantageous here as the evaluation of
$\eta_{\text{N}}= -\partial_t \ln Z_{\text{N}}$ and $\partial_t \lambda$ require up to second-order
  derivatives.\\ 
Introducing the dimensionless Newton coupling and cosmological constant,
\begin{eqnarray}
G
=  \frac{1}{32  \pi \,\bar{\kappa}^2\,  Z_{\text{N}}\, k^{2-d}}\,
&\Rightarrow&\, \partial_t G =  (d-2+\eta_{\text{N}}) G
\nonumber\\
%
\lambda= \bar{\lambda} k^{-2} \,& \Rightarrow & \,\partial_t \lambda = -2
\lambda + k^{-2}\partial_t \bar{\lambda},
\end{eqnarray}
we arrive at an explicit form for the ghost anomalous dimension, which is
given in appendix \ref{etac} for the Landau-deWitt ($\alpha=0$) and the deDonder ($\alpha=1$) gauge.

Endowing the ghost propagator with a non-trivial wave function renormalization
$\eta_c$ feeds back into the flow equations in the Einstein-Hilbert sector by
virtue of the ghost loop. We evaluate the graviton anomalous dimension
$\eta_{\text{N}}$ and the flow of the cosmological constant
$\partial_t \lambda$ by expanding the propagators in a basis of tensor,
vector and scalar hyperspherical harmonics on a $d$ sphere and find the $\beta$ functions given in Appendix \ref{EHbetafunctions}.\\
%
%
This flow of the Einstein-Hilbert sector has been analyzed in a variety of
computations in the literature, e.g. \cite{Reuter:1996cp,Reuter:2001ag, Codello:2008vh,Lauscher:2001rz,Bonanno:2004sy,Fischer:2006fz}. From a technical
viewpoint, our computational method differs from the literature as we do not
rely on heat-kernel results. Apart from this minor detail, our results are in
perfect agreement with the literature; in particular, our choice of a
spectrally adjusted regulator corresponds to type III in
\cite{Codello:2008vh}.  The new modifications arising from the flow of the
ghost sector correspond to the additional terms $\propto \eta_c$.

For the UV fixed-point search, we insert the expression for $\eta_c$ into
\Eqref{etaN} and \Eqref{dtlambda}. Imposing the fixed-point condition $\eta_{\text{N}}
=-2$ and $\partial_t \lambda=0$, we then find the fixed point values and eigenvalues $\theta_{1,2}$ of the stability matrix, which may be compared to their counterparts in a truncation where $\eta_c=0$:
\begin{widetext}
\begin{center}
\begin{figure}[!here]
\begin{tabular}{c|l|c|c|c|c}
 gauge&$\qquad G_{\ast}$&$\lambda_{\ast}$&$G_{\ast}\lambda_{\ast}$&$\theta_{1,2}$&$\eta_c$\\
\hline
Landau-deWitt with $\eta_c =0$ &\, 0.270068 & 0.378519&0.102226&2.10152 $\pm i$1.68512&0\\
Landau-deWitt with $\eta_c \neq0$ &\, 0.28706 & 0.316824 & 0.0909475&2.03409$\pm i$ 1.49895&-0.778944\\
deDonder with $\eta_c=0$&\, 0.181179 &\, 0.480729 &0.0870979&1.40864 $\pm i$ 1.6715&0\\
deDonder with $\eta_c \neq0$&\, 0.207738 &0.348335&0.0723625&1.38757$\pm i$1.283&-1.31245
\end{tabular}
\label{table_FPvalues}
\end{figure}
\end{center}
\end{widetext}
The gauge dependences for the fixed-point values $\lambda_\ast$ and $G_\ast$
agree with those found e.g. in \cite{Lauscher:2001ya} for different regulators. The
value of the ghost anomalous dimension also exhibits a sizable gauge
dependence. Our results provide evidence that $\eta_c <0$ is valid for all
admissible choices of $\alpha$ with a maximum value of $\eta_{c,\text{
    max}}=-0.778944$ being acquired in the Landau-deWitt gauge in the present
truncation. This is due to the fact that the conformal mode contributes with a
positive sign to $\eta_c$ (for $G_{\ast}>0$), whereas the other modes
typically contribute with a negative sign. As the scalar and the vector
contribution are both proportional to $ \alpha$, $\eta_c$ decreases with
increasing $\alpha$.

Another important observation is that the inclusion of the ghost anomalous
dimension on average leads to a weaker gauge dependence in the
Einstein-Hilbert sector, as might be read off from table
\ref{table_FPvalues}. This can be taken as a signature for the convergence of
the expansion of the effective action.

We observe a rather strong dependence of the cosmological-constant fixed point
$\lambda_\ast$ on the ghost anomalous dimension, whereas $G_{\ast}$ shows only
a mild variation. In order to test the stability of the non-Gau\ss ian fixed
point in the Einstein-Hilbert sector against further variations of the ghost
sector, let us study the $\eta_c$ dependence of the fixed point from a more
general viewpoint. For this, we consider $\eta_c$ as a free parameter for the
moment, and then solve the fixed-point equations for $G$ and $\lambda$ as well
as the corresponding critical exponents as a function of $\eta_c$. This can
mimic the influence of higher-order ghost operators on the true value of
$\eta_c$.\\
\begin{figure}[!here]
\begin{center}  
\includegraphics[scale=0.7]{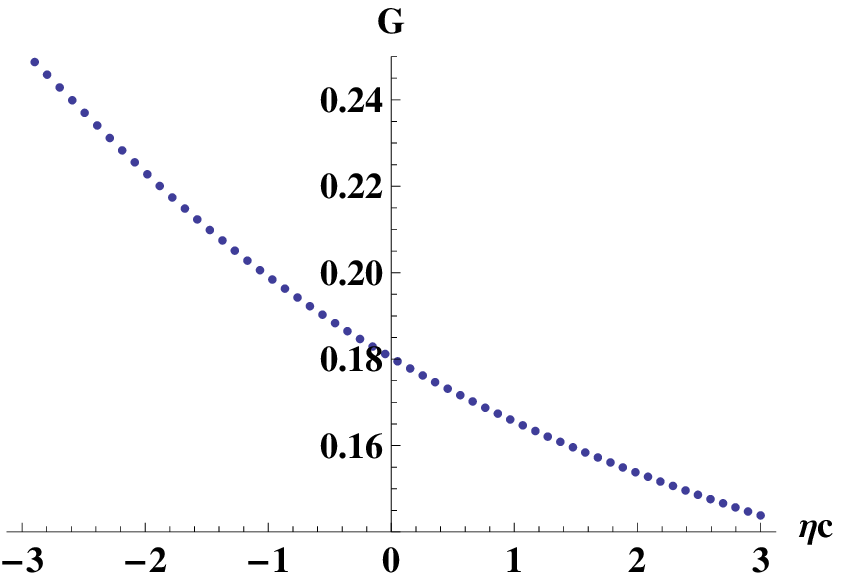}\\
$\phantom{x}$\\
 \includegraphics[scale=0.7]{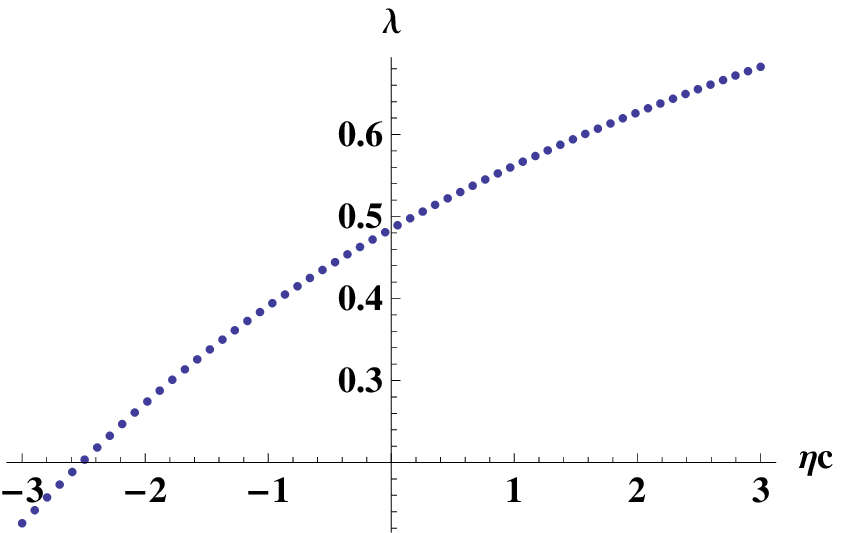}
\end{center}
\caption{Fixed-point values of $G$ (upper panel), $\lambda$ (lower panel) as a function of $\eta_c$ treated as a
  free parameter in the deDonder gauge with $\rho=1=\alpha$. The corresponding self-consistent solution for the ghost anomalous
  dimension is $\eta_c\simeq -1.31245$.}
\end{figure}
\begin{figure}[!here]
\begin{center}
\includegraphics[scale=0.7]{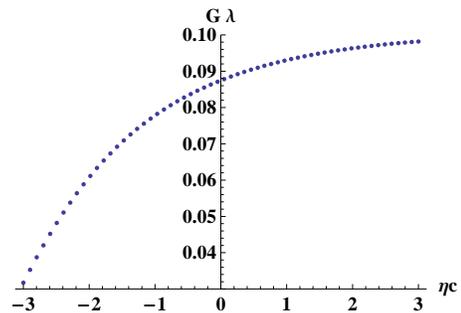}
\end{center}
\caption{Fixed-point value of the product $G \lambda$ as a function of $\eta_c$ treated as a
  free parameter in the deDonder gauge.
}
\end{figure}
\begin{figure}[!here]
\begin{center}  
\includegraphics[scale=0.7]{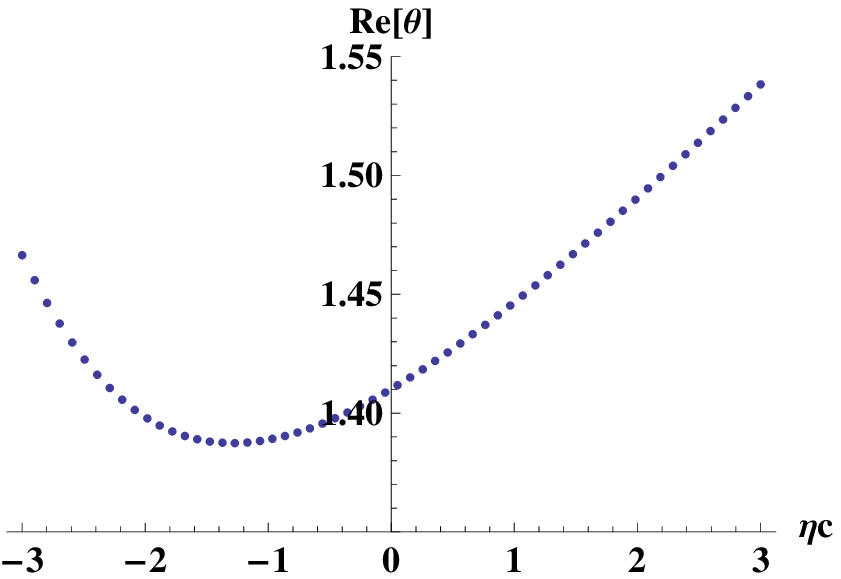}\\
$\phantom{x}$\\
  \includegraphics[scale=0.7]{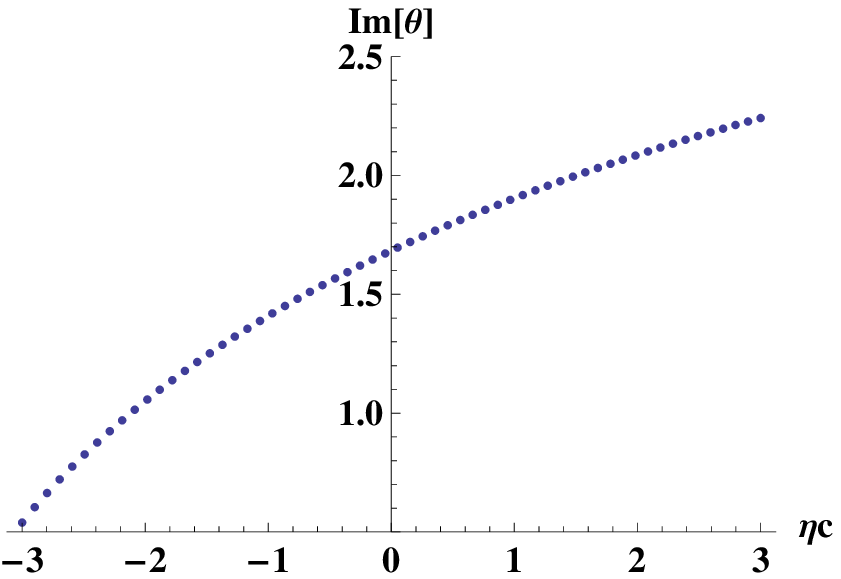}
\end{center} 
\caption{Real part and positive imaginary part of the critical exponents
   (eigenvalues of the stability matrix) in the Einstein-Hilbert sector as a
   function of $\eta_c$ treated as a free parameter in the deDonder
   gauge. The corresponding self-consistent solution for the ghost
     anomalous dimension is $\eta_c\simeq -1.31245$.}
\end{figure}\\
For any $\eta_c \in [-3,3]$, which we expect to cover the physically relevant
range, we find a physically reasonable fixed point which provides further
evidence for the stability of the asymptotic-safety scenario.  Quantitatively,
the cosmological constant gets significantly reduced for decreasing ghost
anomalous dimension. The values of $G_{\ast}$ show the opposite behavior, but
much less pronounced.  The universal product $G_{\ast}\lambda_{\ast}$ becomes
less sensitive to $\eta_c$ for positive $\eta_c$.

Furthermore, let us investigate the ``TT approximation'', where all graviton modes
except for the transverse traceless mode are not allowed to propagate. This
approximation has effectively been adopted in the study of the ghost-curvature
coupling in \cite{Eichhorn:2009ah}.  As on the
one hand the transverse traceless mode is the genuine spin-2 mode
characteristic for gravity, and on the other hand the TT approximation is
technically much easier to deal with, it is interesting to see whether this
"transverse traceless approximation" is able to capture the essential physics,
such as the existence of the fixed point and the number of relevant
directions. In the TT approximation and in the Landau-deWitt gauge, we
find a fixed point with a negative value of $\lambda$ and a positive $G$. As
there are RG trajectories where the negative cosmological constant is acquired
only in the UV, this fixed point is physically admissible. The critical
exponents in this approximation no longer form a complex conjugate pair of
eigenvalues in the Einstein-Hilbert sector, but two positive real eigenvalues
$\theta_1=2.808,\, \theta_2=1.075$. The ghost anomalous dimension is given by
$\eta_c=-0.5515$. We conclude that the existence of the non-Gau\ss ian fixed
point as well as the number of relevant directions is correctly reproduced in
the TT approximation. It is an interesting open question whether this
feature of the transverse traceless approximation is generic and can be
confirmed also in other truncations.

As another approximation of our full result, we can take the
  perturbative limit, i.e., consider the leading contribution in $G$ to
  $\eta_c$. For vanishing cosmological constant, but arbitrary gauge parameter
  $\alpha$ ($=\rho$), we find
\begin{equation}
\eta_c=-\frac{\left(5 \alpha ^3-26 \alpha ^2+5 \alpha +124\right) G}{4 \pi  (\alpha
  -3)^2} + \mathcal{O}(G^2).
\label{eq:PT}
\end{equation}
Again, we observe that $\eta_c$ is negative for all admissible gauge
parameters $\alpha\geq0$, with a singularity at $\alpha=3$ due to incomplete
gauge fixing as discussed above. The Landau-deWitt-gauge limit is
$\eta_c=-\frac{31}{9\pi} G \simeq -1.0964 G$, being in the same range as at
the non-Gau\ss ian fixed point. Of course, this result is non-universal but
scheme dependent in four dimensions, as the power-counting RG critical
dimension is $d=2$.

Let us finally compare our findings to very recent results by Saueressig and
Groh, where $\eta_c=-1.85$ was obtained in the deDonder gauge applying a
cutoff scheme without spectral adjustment \cite{Saueressig:2010}. The quantitative variation is in
the range familiar from typical regulator dependencies already observed in
earlier calculations, see e.g. \cite{Codello:2008vh}. Most importantly, the
qualitative pictures mutually confirm each other.

%
\section{Conclusions}
\label{sec:conc}

We have computed the ghost anomalous dimension in quantum gravity in various
covariant gauges. Within the asymptotic-safety scenario of quantum gravity,
our work represents a further extension of existing approximation schemes for
the construction of renormalizable RG trajectories. Our results provide
further evidence for the existence of a non-Gau\ss ian fixed point which
facilitates a UV-complete construction of quantum gravity. 

Even though the ghost anomalous dimension turns out to be large, i.e., of order
one, its influence on the fixed-point properties in the Einstein-Hilbert
sector remains rather moderate, confirming earlier approximations that have
neglected $\eta_c$. On the other hand, the values that $\eta_c$ can acquire in
standard gauges indicate that ghost contributions can become important in
future analyses.

The anomalous dimension $\eta_c$ parameterizes the scale
dependence of the wave function renormalization $Z_c$. As the functional RG
flow is local in momentum space, we can consider the scale dependence of $Z_c$
as an estimate for the momentum dependence of the fully dressed inverse ghost
propagator:
\begin{equation}
G_{c}(p^2)^{-1} \sim Z_c(k^2=p^2) p^2 \sim (p^2)^{1-
  \frac{\eta_c}{2}}. \label{eq:prop1}
\end{equation}
The corresponding correlator in position space reads
\begin{equation}
G_c (x,x') \sim  \frac{1}{|x-x'|^{2+\eta_c}}.
\end{equation}
As $\eta_c<0$ for all cases studied in this work, the propagation of ghost
modes is suppressed in the UV compared to the perturbative propagator. For
those gauges (e.g., $\alpha=0$ and $\alpha=1$) where $-2<\eta_c<0$, the dressed
correlator still develops a singularity at short distances. In general, the UV
behavior of higher-order diagrams involving ghosts is improved for $\eta_c<0$
compared to a perturbative framework.

It is interesting to compare this to the graviton anomalous dimension which is
$\eta_{\text{N}}=-2$ at the fixed point, implying an even weaker UV propagation,
$G_{h}(x,x') \sim \ln |x-x'|$ \cite{Lauscher:2005xz}. At first sight, one might expect that a complete
calculation of the ghost anomalous dimension should also result in $\eta_c=-2$
in order to guarantee a cancellation of gauge modes and ghost modes near the
fixed point. However, this cancellation could also come about for differing
anomalous dimensions if this difference is balanced by a corresponding running
of the vertices with ghost legs. A mechanism of this type has for instance
been observed at a non-Gau\ss ian IR fixed point in Landau-gauge Yang Mills
theory \cite{Fischer:2006vf}. On the other hand, if this difference in the
propagation of ghost and graviton modes persisted (say, if the vertices did not
differ much from a trivial scaling), $\eta_c>\eta_{\text{N}}$ would indicate a
dominance of ghost modes for the RG running. This could be a signature of the
dominance of field configurations near the Gribov horizon along the lines of
the Gribov-Zwanziger scenario in gauge theories \cite{Gribov:1977wm}. We
emphasize that a complete discussion of this issue also requires to
distinguish between the background-field anomalous dimension $\eta_{\text{N}}$ and the
corresponding fluctuation-field anomalous dimension. 

Another consequence of $\eta_c<0$ can be read off from the induced scaling of
the vertices involving ghost modes. Consider a vertex with $m$ ghost and $m$
antighost legs parameterized by a coupling $\bar g^{(m)}$. After renormalizing the
ghost fields with $Z_c^{1/2}$, the $\beta$ function for the renormalized
coupling $g^{(m)}\sim \bar g^{(m)}/Z_c^m$ receives a contribution of the type
\begin{equation}
\partial_t g^{(m)}= m \eta_c + \dots.
\end{equation}
For $\eta_c<0$, this increases the RG relevance of this vertex in comparison
to standard power-counting. A simple example is provided by the ghost-curvature
coupling $\sim \bar\zeta \bar{c}^\mu R c_\mu$ considered in
\cite{Eichhorn:2009ah} ($m=1$ in this case). In a TT approximation, the
renormalized coupling has turned out to be RG relevant for $\eta_c=0$. This
conclusion now remains unchanged, as our values $\eta_c<0$ even enhance the RG
relevance of this operator.

\acknowledgments

The authors would like to thank F.~Saueressig, K.~Groh, M.~Reuter and J.M.~Pawlowski for helpful discussions. This work was supported by the DFG
under contract No. Gi 328/5-1 (Heisenberg program) and GK 1523/1.

\begin{appendix}

\section{Vanishing of the tadpole diagram}\label{tadpole}

The vanishing of the graviton tadpole contribution to the running of $Z_c$
can be shown by making use of the second variation of the Christoffel symbol,
\begin{eqnarray}
 \delta^2 \Gamma^{\lambda}_{\sigma \nu}&=& 
 \delta \left(\delta \Gamma^{\lambda}_{\sigma \nu} \right)\nonumber\\
&=& \delta \left(\frac{1}{2}\gamma^{\lambda \tau}
\left( D_{\sigma}h_{\tau \nu}+ D_{\nu}h_{\sigma \tau}- D_{\tau h_{\sigma
    \nu}}\right) \right)
\nonumber\\
&=&
 - h^{\lambda \tau}\left(D_{\sigma}h_{\tau \nu}+ D_{\nu}h_{\sigma \tau}-
 D_{\tau h_{\sigma \nu}} \right)
.\label{Christoffel_var}
\end{eqnarray}
Varying the ghost kinetic term twice with respect to the metric produces the
following type of terms:
\begin{eqnarray}
 \delta^2 (\gamma_{\kappa \nu}D_{\rho}c^{\nu})&=& 
\delta^2 (\gamma_{\kappa \nu}\Gamma_{\rho \lambda}^{\nu})c^{\lambda}\nonumber\\
&=& (2 h_{\kappa \nu} \delta \Gamma^{\nu}_{\rho \lambda}
+ \gamma_{\kappa \nu}\delta^2 \Gamma^{\nu}_{\rho \lambda})c^{\lambda}.
\end{eqnarray}
Inserting the first and second variation of the Christoffel symbol from
\Eqref{Christoffel_var} leads to a cancellation between the two
terms. Accordingly, the second variation of the ghost kinetic term with
respect to the metric vanishes (for all choices of $\rho$). Hence, there is
no gauge in which a graviton tadpole can contribute to the running of the
ghost wave function renormalization. (Clearly a more general kinetic
  ghost term with the volume element $d^4x \sqrt{\gamma}$ would result in a
  tadpole contribution from the second variation of the volume
  element.)

\section{Vertices and propagators}\label{vertices}

In this appendix, we derive the building blocks for the expansion
\eqref{eq:flowexp} of the flow equation in terms of the quantities $\mathcal
P$ and $\mathcal F$. In the following, we always aim at a Euclidean flat
  background. Our conventions for 2-point functions are given by
\begin{equation}
\Gamma_{k,ij}^{(2)}(p,q)= \frac{\overset{\rightarrow}{\delta}}{\delta \phi_i(-p)} 
\Gamma_{k} \frac{\overset{\leftarrow}\delta}{\delta \phi_j(q)},
\end{equation}
where $\phi(p)= \left(h_{\mu \nu}^{\text{T}}(p), v_{\mu}(p), \sigma(p), h(p), c_{\mu}(p),
\bar{c}_{\mu}(-p) \right)$ and $i,j$ label the field components. Here, we have
chosen the momentum-space conventions for the anti-ghost opposite to those of
the ghost, i.e., if $c^\mu(p)$ denotes a ghost with {\em incoming} momentum
$p$ then $\bar{c}^\mu(q)$ denotes an anti-ghost with {\em outgoing} momentum
$q$. The ghost propagator is an off-diagonal matrix,
\begin{eqnarray}
\mathcal{P}_{\text{gh}}^{-1}&=&\left( \begin{array}{cc}
                 0& \Gamma^{(2)}_{k, c \bar{c}}(p,q)+R_k\\
 \Gamma^{(2)}_{k,\bar{c} c }(p,q))+R_k & 0
                    \end{array} \right)^{-1}\\
                    &=&\left( \begin{array}{cc}
                 0& \left( \Gamma^{(2)}_{k,\bar{c} c}(p,q)+R_k\right)^{-1}\\
\left( \Gamma^{(2)}_{k,c \bar{c}}(p,q)+R_k\right)^{-1}&0
                    \end{array} \right)\nonumber\\
&=&\left( \begin{array}{cc}
                 0& \mathcal{P}_{c\bar c}^{-1}\\
\mathcal{P}_{\bar c c}^{-1}&0
                    \end{array} \right),
\end{eqnarray}
where
\begin{eqnarray}
 \Gamma^{(2)}_{k,c \bar{c} , \mu \nu}(p,q)&=&
 \frac{\overset{\rightarrow}{\delta}}{\delta c_{\mu}(-p)}\Gamma_{k\, \rm
   gh}\frac{\overset{\leftarrow}{\delta}}{\delta \bar{c}_{\nu}(-q)}= -
 \Gamma^{(2)}_{k,\bar{c} c , \mu \nu}(p,q)).\nonumber\\
&{}& 
\end{eqnarray}
Within our truncation, the ghost propagator reads explicitly
\begin{eqnarray}
&{}&\left(\Gamma_{k,\bar{c} c}^{(2)}+R_k\right)_{\mu \nu}^{-1}\\
&=& \frac{1}{\sqrt{2}Z_c p^2}\left(\delta_{\mu \nu}+ \frac{\rho-1}{3-\rho}\frac{p_{\mu}p_{\nu}}{p^2}\right)\frac{1}{(1+r(y))}\delta^{4}(p-q).\nonumber
\end{eqnarray}
The graviton propagators are obtained from the second variation of the
Einstein-Hilbert and the gauge-fixing action. Setting $\gamma_{\mu
  \nu}=\bar{g}_{\mu \nu}=\delta_{\mu \nu}$ after the functional variation yields the following
expression in Fourier space:
\begin{eqnarray}
&{}&\delta^2\Gamma_{k\, \rm EH + gf}\nonumber\\
&=& \bar{\kappa}^2 \int \frac{d^4p}{(2\pi)^4} h^{\alpha \beta}(-p)\Bigl[\frac{1}{4}\left(\delta_{\alpha \mu}\delta_{\beta \nu} +\delta_{\alpha \nu}\delta_{\beta \mu}\right)p^2\nonumber\\
&-&\frac{1}{2}\delta_{\alpha \beta}p^2 \delta_{\mu \nu} + \delta_{\alpha \beta}p_{\mu}p_{\nu} -\frac{1}{2}\left(p_{\beta}p_{\mu}\delta_{\nu \alpha}+p_{\alpha}p_{\nu}\delta_{\mu \beta} \right) \nonumber\\
&+&\lambda_k \left(\frac{1}{2}\delta_{\alpha \beta}\delta_{\mu \nu} -\frac{1}{2}\left(\delta_{\alpha \mu}\delta_{\beta \nu}+\delta_{\alpha \nu} \delta_{\beta \mu} \right) \right)\nonumber\\
&+&\frac{1}{\alpha}\Bigl(\frac{1}{2}\left(p_{\alpha}p_{\mu}\delta_{\beta \nu}+ p_{\beta}p_{\mu} \delta_{\beta \mu}\right)-\frac{1+\rho}{2}p_{\alpha}p_{\beta}\delta_{\mu \nu}\nonumber\\
&+&\phantom{\frac{1}{\alpha} \Bigl(}\frac{(1+\rho)^2}{16} \delta_{\alpha \beta}p^2 \delta_{\mu \nu}\Bigr)  \Bigr]h_{\mu \nu}(p).\label{Gamma2EHgf}
\end{eqnarray}
Inserting the York decomposition \eqref{York} into \Eqref{Gamma2EHgf} then
results in the following expression, from which the inverse propagators follow
directly by functional derivatives:
\begin{widetext}
\begin{eqnarray}
\delta^2\Gamma_{k\, \rm EH + gf}&=& Z_\text{N} \bar{\kappa}^2 \int
\frac{d^4p}{(2\pi)^4}
\Bigl[h^{\text{T}\,\alpha \beta} (-p) \frac{1}{2}\left(p^2-2\lambda_k \right)
h_{\alpha \beta}^T(-p)+ v^{\beta}(-p)\left(\frac{p^2}{\alpha}-2 \lambda_k
\right)v_{\beta}(p)\nonumber\\
&+& \sigma(-p)\frac{3}{16}\left(p^2 \frac{3-\alpha}{\alpha}-4 \lambda_k
\right)\sigma(p)
+ h(-p)3\frac{\rho-\alpha}{8 \alpha}p^2\sigma(-p)+\frac{1}{16}h(-p)\left( p^2 \frac{\rho^2 -3 \alpha}{\alpha}+4 \lambda
\right)h(p) \Bigr]. \label{Gamma2EHYork}
\end{eqnarray}
\end{widetext}
In this work, we confine ourselves to the gauge choice $\rho \rightarrow
\alpha$ where the propagator matrix becomes diagonal in the graviton modes. The
vector and transverse traceless tensor propagators go along with transverse and
transverse traceless projectors, respectively. In $d$-dimensional Fourier
space, these projectors read
\begin{eqnarray}
 {P}_{\text T\, \mu \nu}(p)&=& \delta_{\mu \nu} -
 \frac{p_{\mu}p_{\nu}}{p^2},\nonumber\\ 
{P}_{\text{TT}\, \mu \nu \kappa \lambda}(p)&=&
\frac{1}{2}\left({P}_{\text T\, \mu \kappa}{P}_{\text T\, \nu \lambda}+
  {P}_{\text T\, \mu \lambda}{P}_{\text T\, \nu \kappa}\right)\nonumber\\
&&-
\frac{1}{d-1}{P}_{\text T\, \mu \nu} {P}_{\text T\, \kappa \lambda}, 
\end{eqnarray}
where the last term in the transverse traceless projector ${P}_{\text{TT}}$
removes the trace part.

The resulting propagators together with the regulator $R_k$ constitute the
$\mathcal P$ term in the expansion of the flow equation
\eqref{eq:flowexp}.

 The $\mathcal F$ term carries the dependence on the ghost
fields that couple via vertices to the fluctuation modes. 
To obtain these vertices, we vary the ghost action once with respect to
the metric and then proceed to a flat background,
yielding:
\begin{widetext}
\begin{eqnarray}
&{}& \delta \Gamma_{k \, \rm gh}\nonumber\\
&=& -\sqrt{2} Z_c \int d^4x \sqrt{\bg} 
\bar{c}^{\mu} \Bigl(\bar{D}^{\rho}h_{\mu \nu}\bar{D}_{\rho}
+ \bar{D}^{\rho}\left[\bar{D}_{\nu}h_{\mu \rho}\right] 
+\bar{D}^{\rho}h_{\rho \nu}\bar{D}_{\mu}
- \frac{1}{2}(1+\rho)\bar{D}_{\mu}h^{\rho}_{\nu}\bar{D}_{\rho}-\frac{1}{4}(1+\rho)\bar{D}_{\mu}[\bar{D}_{\nu}h_{\lambda}^{\, \lambda}]
\Bigr)c^{\nu}.\nonumber\\ 
&\rightarrow& -\sqrt{2}Z_c \int \frac{d^4p}{(2 \pi)^4}\frac{d^4 q}{(2 \pi)^4}
 \bar{c}^{\mu}(p+q)h_{\rho \sigma}(p)c^{\kappa}(q)
\Bigl(-q\cdot(p+q)\frac{1}{2}\left(\delta_{\mu}^{\rho}\delta_{\kappa}^{\sigma}
+\delta^{\sigma}_{\mu}\delta^{\rho}_{\kappa}\right)-\frac{p_{\kappa}}{2}\left((p^{\rho}+q^{\rho})\delta^{\sigma}_{\mu}
+(p^{\sigma}+q^{\sigma})\delta_{\mu}^{\rho}\right)\nonumber\\
&{}&-\frac{q_{\mu}}{2}\left((p^{\sigma}
+q^{\sigma})\delta_{\kappa}^{\rho}+(p^{\rho}+q^{\rho})\delta^{\sigma}_{\kappa}\right)+\frac{1}{2}\frac{1+\rho}{2}(p_{\mu}+q_{\mu})
\left(\delta_{\kappa}^{\rho}q^{\sigma}+\delta^{\sigma}_{\kappa}q^{\rho}\right)
+\frac{1+\rho}{4}p_{\kappa}\delta^{\rho \sigma}(p_{\mu}+q_{\mu})\Bigr)\nonumber\\
%
&=&-\sqrt{2}Z_c \int \frac{d^4p}{(2 \pi)^4}\frac{d^4 q}{(2 \pi)^4}
\bar{c}^{\mu}(p+q)c^{\kappa}(q) \Bigl(V^{(\text{T})}_{\kappa \mu}{}^{\rho
  \sigma}(p,q)h_{\rho \sigma}^{\text T}(p)
+V^{(v)}_{\kappa \mu}{}^{\rho}(p,q)v_{\rho}(p)+
+V^{(\sigma)}_{\kappa \mu}(p,q)\sigma(p)        
+V^{(h)}_{\kappa \mu}(p,q)h(p) \Bigr).\nonumber
\end{eqnarray}
\end{widetext}
Here, we introduced the York decomposition \eqref{York} for the graviton
fluctuation, such that we can read off the corresponding vertices connecting
ghost and anti-ghost with the graviton components:
\begin{widetext}
\begin{eqnarray}
 V^{(\text{T})}_{\kappa \mu}{}^{\rho \sigma }(p,q)&=& -q\cdot(p+q)\frac{1}{2}
\left(\delta_{\mu}^{\rho}\delta_{\kappa}^{\sigma}
+\delta^{\sigma}_{\mu}\delta^{\rho}_{\kappa}\right)
-\frac{p_{\kappa}}{2}\left(q^{\rho}\delta^{\sigma}_{\mu}
+q^{\sigma}\delta_{\mu}^{\rho}\right)-\frac{q_{\mu}}{2}\left(q^{\sigma}
\delta_{\kappa}^{\rho}+q^{\rho}\delta^{\sigma}_{\kappa}\right)+\frac{1}{2}\frac{1+\rho}{2}(p_{\mu}+q_{\mu})
\left(\delta_{\kappa}^{\rho}q^{\sigma}+\delta^{\sigma}_{\kappa}q^{\rho}\right),\nonumber\\
V^{(v)}_{\kappa \mu}{}^{\rho }(p,q)&=& \frac{2 i}{\sqrt{p^2}} p_{\sigma}
\Bigl(-q\cdot(p+q)\frac{1}{2}\left(\delta_{\mu}^{\rho}\delta_{\kappa}^{\sigma}
+\delta^{\sigma}_{\mu}\delta^{\rho}_{\kappa}\right)
-\frac{p_{\kappa}}{2}\left(q^{\rho}\delta^{\sigma}_{\mu}
+(p^{\sigma}+q^{\sigma})\delta_{\mu}^{\rho}\right)-\frac{q_{\mu}}{2}\left((p^{\sigma}+q^{\sigma})\delta_{\kappa}^{\rho}
+q^{\rho}\delta^{\sigma}_{\kappa}\right)\nonumber\\
&{}&+\frac{1}{2}\frac{1+\rho}{2}(p_{\mu}
+q_{\mu})\left(\delta_{\kappa}^{\rho}q^{\sigma}+\delta^{\sigma}_{\kappa}
q^{\rho}\right) \Bigr) \label{V_terms}\\
V^{(\sigma)}_{\kappa \mu}(p,q)&=&- \frac{1}{p^2}\Bigl[p_{\kappa}p_{\mu}
\left(\frac{3}{4}p^2+q^2+\frac{3-\rho}{2}q\cdot p \right)
+q_{\mu}q_{\kappa}\frac{1}{4}p^2\left(\frac{\rho-1}{2} \right)-\frac{1}{4}p^2\delta_{\mu \kappa}\left(q^2+q\cdot p \right)\nonumber\\
&{}&+p_{\kappa}q_{\mu}\left(\frac{1}{2}p^2+\frac{1-\rho}{2}q\cdot p \right)
+p_{\mu}q_{\kappa}\frac{1+\rho}{8}p^2\Bigr]\nonumber\\
V^{(h)}_{\kappa \mu}(p,q)&=&\frac{1}{4}\Bigl(-q\cdot(p+q)
\delta_{\mu \kappa}+p_{\kappa}(p_{\mu}+q_{\mu})(\rho -1) 
+ p_{\mu}p_{\kappa}-q_{\mu}q_{\kappa}\frac{1-\rho}{2}
+p_{\mu}q_{\kappa}\frac{1+\rho}{2}\Bigr).\nonumber
\end{eqnarray}
\end{widetext}
From this, the four possible fluctuation matrix entries contributing to the
quantity $\mathcal F$ in the expansion \eqref{eq:flowexp} can be evaluated:\\
\begin{eqnarray}
\Gamma^{(2)}_{h^{\text{T}} c}(q,p)&=& 
\frac{\overset{\rightarrow}{\delta}}{\delta h^{\text{T}}_{\mu \nu}(-q)}
\Gamma_{k\, \rm gh}\frac{\overset{\leftarrow}{\delta}}{\delta c_{\kappa}(p)}\nonumber\\
&=& -\sqrt{2}Z_c\bar{c}^{\tau}(p-q)V_{\kappa \tau \mu \nu}^{(\text{T})}(-q,p)\nonumber\\
\Gamma^{(2)}_{h^{\text{T}} \bar{c}}(q,p)&=& 
\frac{\overset{\rightarrow}{\delta}}{\delta h^{\text{T}}_{\mu \nu}(-q)}
\Gamma_{k\, \rm gh}\frac{\overset{\leftarrow}{\delta}}{\delta
  \bar{c}_{\kappa}(-p)}
\nonumber\\
&=&\sqrt{2} Z_c c^{\tau}(q-p)V^{(\text{T})}_{\tau \kappa \mu \nu}(-q,q-p)\nonumber\\
\Gamma^{(2)}_{c h^{\text{T}}}(q,p)&=& 
\frac{\overset{\rightarrow}{\delta}}{\delta c_{\kappa}(-q)}
\Gamma_{k\, \rm gh}\frac{\overset{\leftarrow}{\delta}}{\delta
  h^{\text{T}}_{\mu \nu}(p)}
\nonumber\\ 
&=& \sqrt{2}Z_c \bar{c}^{\lambda}(p-q)V^{(\text{T})}_{\kappa \lambda \mu
  \nu}(p,-q)
\nonumber\\
\Gamma^{(2)}_{\bar{c} h^{\text{T}}}(q,p)&=& 
\frac{\overset{\rightarrow}{\delta}}{\delta \bar{c}_{\kappa}(q)}
\Gamma_{k\, \rm gh}\frac{\overset{\leftarrow}{\delta}}{\delta h^{\text{T}}_{\mu \nu}(p)}\nonumber\\
&=& -\sqrt{2}Z_c c^{\lambda}(q-p)V^{(\text T)}_{\lambda \kappa \mu \nu}(p,q-p),
\end{eqnarray}
and similarly for the other graviton modes.


%

\section{$\beta$ functions in the Einstein-Hilbert sector}\label{EHbetafunctions}
In the Einstein-Hilbert sector, we find the following $\beta$ functions,
obtained with a spectrally and RG adjusted regulator with exponential shape
function in $d=4$:
\begin{widetext}
\begin{eqnarray}
&{}&\partial_t \lambda+ 2\lambda \nonumber\\
&=&\frac{1}{36\pi} \Biggl(G \lambda \Biggl\{-150 (\partial_t\lambda+2 \lambda ) \text{Ei}(2 \lambda )-\frac{12 (\partial_t\lambda+2 \lambda ) \text{Ei}\left(-\frac{4 \lambda }{\alpha -3}\right)}{\alpha -3}-150
   \text{Li}_2\left(e^{2 \lambda }\right)+6 \text{Li}_2\left(e^{-\frac{4 \lambda }{\alpha -3}}\right)+36 \text{Li}_2\left(e^{2 \alpha  \lambda }\right)\nonumber\\
&+&\frac{6}{(\alpha -3)^2} \Biggl(-3 \alpha  (3 \alpha -2) (\partial_t\lambda+2 \lambda ) \text{Ei}(2 \alpha  \lambda ) (\alpha -3)^2-9 \alpha  \text{Li}_2\left(e^{2 \alpha  \lambda }\right) (\alpha -3)^2+(7 \alpha -9) \Bigl[\text{Li}_2\left(e^{-\frac{4 \alpha  \lambda }{\alpha
   -3}}\right) (\alpha -3)\nonumber\\
&-&2 \alpha  (\partial_t\lambda+2 \lambda ) \text{Ei}\left(-\frac{4 \alpha  \lambda }{\alpha -3}\right)\Bigr]-2 i \pi\lambda  \Bigl[\alpha  (39 \alpha -166)+219\Bigr]+\partial_t\lambda \Biggl[25 \ln \left(-1+e^{2 \lambda }\right) (\alpha -3)^2\nonumber\\
&+&2 \ln \left(-1+e^{-\frac{4 \lambda }{\alpha -3}}\right) (\alpha -3)+\alpha  \Big[3 (3 \alpha -2) \ln \left(-1+e^{2 \alpha  \lambda }\right)
   (\alpha -3)^2+2 (7 \alpha -9) \ln \left(-1+e^{-\frac{4 \alpha  \lambda }{\alpha -3}}\right)\Bigr]\Biggr]\Biggr)\nonumber\\
&+&\frac{1}{\alpha -3}e^{-\frac{4 (\alpha +1) \lambda }{\alpha -3}} \Biggl(-2 e^{\frac{4
   (\alpha +1) \lambda }{\alpha -3}} \Bigl(18 i \pi  (\alpha -3) \alpha  (3 \alpha -2) \lambda+(7 \alpha -27) \left(\pi ^2-3 \eta_c\right)\Bigr)-3 e^{\frac{4 \alpha  \lambda }{\alpha -3}} (\alpha -3) \eta_{\text{N}}\nonumber\\
&+&75 e^{\frac{2 (3 \alpha -1) \lambda }{\alpha -3}} (\alpha -3) \eta_{\text{N}}+9 e^{\frac{2 ((\alpha -1) \alpha +2) \lambda }{\alpha -3}} (\alpha -3) (3 \alpha -2) \eta_{\text{N}}-3 e^{\frac{4 \lambda }{\alpha -3}} (7
   \alpha -9) \eta_{\text{N}}\Biggr)\Biggr\}\nonumber\\
& +& 9 G \Biggl[\frac{8 \alpha  \lambda  \text{Li}_2\left(e^{-\frac{4 \alpha  \lambda }{\alpha -3}}\right)}{\alpha -3}+\frac{4\alpha  e^{\frac{-4 \alpha  \lambda }{\alpha -3}}}{(\alpha -3)^2}
   (\partial_t\lambda+2 \lambda ) \Biggl\{\alpha +e^{\frac{4 \alpha  \lambda }{\alpha -3}} \Biggl(4 \alpha  \lambda  \left(\text{Ei}\left(\frac{-4 \alpha 
   \lambda }{\alpha -3}\right)+i \pi \right)-(\alpha -3) \text{Li}_2\Bigl[e^{\frac{-4 \alpha  \lambda }{\alpha -3}}\Bigr]\Biggr)-3\Biggr\}\nonumber\\
&+&6 \Biggl[2 \lambda  (\partial_t\lambda+2 \lambda )
   (\text{Ei}(2 \alpha  \lambda )+i \pi ) \alpha ^2+\partial_t\lambda \text{Li}_2\left(e^{2 \alpha  \lambda }\right) \alpha +2 \text{Li}_3\left(e^{2 \alpha  \lambda }\right)\Biggr]+4
   \text{Li}_3\left(e^{-\frac{4 \alpha  \lambda }{\alpha -3}}\right)-e^{-\frac{4 \alpha  \lambda }{\alpha -3}} \eta_{\text{N}}\nonumber\\
&-&3 e^{2 \alpha  \lambda } \left(2 \partial_t\lambda \alpha +4 \lambda  \alpha +\eta
   _N\right)+5 \Biggl[4 \lambda  (\partial_t\lambda+2 \lambda ) (\text{Ei}(2 \lambda )+i \pi )+2\partial_t\lambda \text{Li}_2\left(e^{2 \lambda }\right)+4 \text{Li}_3\left(e^{2 \lambda }\right)-e^{2 \lambda
   } \left(2 \partial_t\lambda+4 \lambda +\eta_{\text{N}}\right)\Biggr]\nonumber\\
&+&\frac{4}{(\alpha -3)^2}\Biggl( \text{Li}_3\left(e^{-\frac{4 \lambda }{\alpha -3}}\right) (\alpha -3)^2-4 \partial_t\lambda \text{Li}_2\left(e^{-\frac{4 \lambda
   }{\alpha -3}}\right) (\alpha -3)+e^{-\frac{4 \lambda }{\alpha -3}} \left(4 \partial_t\lambda+8 \lambda -(\alpha -3) \eta_{\text{N}}\right) (\alpha -3)\nonumber\\
&+&16 i \pi  \lambda  (\partial_t\lambda+2 \lambda )+16 \lambda 
   (\partial_t\lambda+2 \lambda ) \text{Ei}\left(-\frac{4 \lambda }{\alpha -3}\right)\Biggr)+8 \left(\eta_c-4 \zeta (3)\right)\Biggr]\Biggr), \quad \mbox{ for $\lambda>0$.}\label{dtlambda}
\end{eqnarray}
\begin{eqnarray}
\eta_{\text{N}}&=&\frac{G}{36 \pi } \Biggl[54 e^{2 \alpha  \lambda } (\partial_t \lambda+2 \lambda ) \alpha ^2-\frac{144 \lambda  \ln \left(1-e^{-\frac{4 \alpha  \lambda }{\alpha -3}}\right) \alpha ^2}{(\alpha -3)^2}-36 e^{2 \alpha 
   \lambda } (\partial_t \lambda+2 \lambda ) \alpha-18 (3 \alpha -2) (\partial_t \lambda+2 \lambda ) \text{Ei}(2 \alpha  \lambda ) \alpha \nonumber\\
&-&18 (3 \alpha -2) \left(2 i \pi  \lambda +e^{2 \alpha  \lambda }
   (\partial_t \lambda+2 \lambda )-\partial_t \lambda \ln \left(-1+e^{2 \alpha  \lambda }\right)\right) \alpha-\frac{24 \lambda  \ln \left(1-e^{-\frac{4 \alpha  \lambda }{\alpha -3}}\right) \alpha }{\alpha
   -3}+\frac{144\lambda }{(\alpha -3)^2}  \ln \left(1-e^{-\frac{4 \alpha  \lambda }{\alpha -3}}\right) \alpha\nonumber\\
&-&\frac{12 (\partial_t \lambda+2 \lambda ) \text{Ei}\left(-\frac{4 \lambda }{\alpha -3}\right)}{\alpha -3}+\frac{12 (7 \alpha -9)}{(\alpha -3)^2} (\partial_t \lambda+2 \lambda ) \left(\ln \left(-1+e^{-\frac{4 \alpha 
   \lambda }{\alpha -3}}\right)-\text{Ei}\left(-\frac{4 \alpha  \lambda }{\alpha -3}\right)\right) \alpha +27 e^{2 \alpha  \lambda } \eta_{\text{N}} \alpha\nonumber\\
&-&\frac{3 e^{-\frac{4 \lambda }{\alpha -3}}
   \eta_{\text{N}} \alpha }{\alpha -3}-150 (\partial_t \lambda+2 \lambda ) \text{Ei}(2 \lambda )-300 \lambda 
   \ln \left(1-e^{2 \lambda }\right)+150 \partial_t \lambda \ln \left(-1+e^{2 \lambda }\right)+300 \lambda  \ln \left(-1+e^{2 \lambda }\right)\nonumber\\
&-&\frac{24 \lambda  \ln \left(1-e^{-\frac{4 \lambda }{\alpha
   -3}}\right)}{\alpha -3}+\frac{24 \lambda  \ln \left(-1+e^{-\frac{4 \lambda }{\alpha -3}}\right)}{\alpha -3}+\frac{12 \partial_t \lambda \ln \left(-1+e^{-\frac{4 \lambda }{\alpha -3}}\right)}{\alpha
   -3}-150 \text{Li}_2\left(e^{2 \lambda }\right)+6 \text{Li}_2\left(e^{-\frac{4 \lambda }{\alpha -3}}\right)\nonumber\\
&+&18 (2-3 \alpha ) \text{Li}_2\left(e^{2 \alpha  \lambda }\right)+\frac{6 (7 \alpha -9)
   \text{Li}_2\left(e^{-\frac{4 \alpha  \lambda }{\alpha -3}}\right)}{\alpha -3}-\frac{2 (\alpha -9) \left(\pi ^2-3 \eta_c\right)}{\alpha -3}-12 \left(\pi ^2-3 \eta_c\right)+75 e^{2 \lambda } \eta_{\text{N}}-18 e^{2
   \alpha  \lambda } \eta_{\text{N}}\nonumber\\
&+&\frac{9 e^{-\frac{4 \lambda }{\alpha -3}} \eta_{\text{N}}}{\alpha -3}+\frac{3 e^{-\frac{4 \alpha  \lambda }{\alpha -3}} (9-7 \alpha ) \eta_{\text{N}}}{\alpha -3}\Biggr]\nonumber\\
&{}&\mbox{ for $\lambda>0$.} \label{etaN}
\end{eqnarray}
\end{widetext}
\section{Ghost anomalous dimension}\label{etac}
In the Landau-deWitt gauge ($\alpha=0$), we find the following expression for
$\eta_c$, where the three lines are the transverse traceless contribution
and the last lines are due to the conformal mode:
\begin{widetext}
\begin{eqnarray}
\eta_{c\, \text{L}}
&=& -\frac{35}{162 \pi } e^{-4 \lambda } G \Biggl(e^{6 \lambda } \bigr[6 \partial_t\lambda+\eta_c-3 \eta_{\text{N}}+6\bigl]+e^{8 \lambda } \eta_{\text{N}}+3 e^{4 \lambda } \bigl[-\eta_c+8 \lambda +4\bigr]+12 \Bigl(e^{2
   \lambda } \bigl[\partial_t\lambda (6 \lambda -1)\nonumber\\
&+&\lambda  (-\eta_{\text{N}}+12 \lambda -2)\bigr]-\eta_c \lambda \Bigr) E_1(-6 \lambda )+12 e^{2 \lambda } \Bigl(\partial_t\lambda e^{2 \lambda } (4
   \lambda -1)+\lambda  \Bigl[-\eta_c+4 \lambda +e^{2 \lambda } (-\eta_{\text{N}}+8 \lambda -2)\Bigr]\Bigr) E_1(-2 \lambda )\nonumber\\
&+&12 \Bigl(\eta_c \lambda +e^{2 \lambda } \Bigl(-6 \lambda 
   \partial_t\lambda+\partial_t\lambda+(\eta_c+\eta_{\text{N}}-16 \lambda +2) \lambda \Bigr)+e^{4 \lambda } (-4 \lambda  \partial_t\lambda+\partial_t\lambda+\eta_{\text{N}}-8 \lambda +2) \lambda
   \Bigr) \Gamma (0,-4 \lambda )\Biggr)\nonumber\\
&+&\frac{2}{243 \pi } e^{-8 \lambda /3} G \Biggl(-8 e^{4 \lambda /3} \Bigl(\lambda  (8 \lambda -3 \eta_c)+e^{4 \lambda /3} \bigr[\partial_t\lambda (8 \lambda -3)+\lambda  (-3 \eta_{\text{N}}+16 \lambda -6)\bigl]\Bigr)
   \text{Ei}\left(\frac{4 \lambda }{3}\right)\nonumber\\
&+&8 \Bigl[-3 \eta_c \lambda +e^{8 \lambda /3} \bigl[\partial_t\lambda (8 \lambda -3)+\lambda  (-3 \eta_{\text{N}}+16 \lambda -6)\bigr]+e^{4 \lambda /3} \bigl[3
  \partial_t\lambda (4 \lambda -1)+\lambda  (32 \lambda -3 (\eta_c+\eta_{\text{N}}+2))\bigr])\Bigr] \text{Ei}\left(\frac{8 \lambda }{3}\right)\nonumber\\
&+&3 \Biggl[e^{8 \lambda /3} \Bigl(-3 \eta_c+e^{4
   \lambda /3} \Bigl(4 \partial_t\lambda+\eta_c+\left(-3+e^{4 \lambda /3}\right) \eta_{\text{N}}+6\Bigr)+16 \lambda +12\Bigr)\nonumber\\
&+&8 \left(\eta_c \lambda +e^{4 \lambda /3} (-4 \lambda 
   \partial_t\lambda+\partial_t\lambda+(\eta_{\text{N}}-8 \lambda +2) \lambda )\right) \text{Ei}(4 \lambda )\Biggr]\Biggr).
\label{etac_Landau}
\end{eqnarray}
\end{widetext}
The exponential factors result from the spectral adjustment of the
regulator. The expression is linear in $G$, as each contributing diagram contains
exactly one graviton propagator.

In the deDonder or harmonic gauge ($\alpha=1$), we have to take contributions
from all modes into account. Accordingly, we arrive at the following
expression, which decomposes into transverse traceless \eqref{etacTT}, vector
\eqref{etacv}, scalar \eqref{etacsc} and conformal \eqref{etacphi}
contributions:
\begin{widetext}
\begin{eqnarray}
\eta_{c\, \text{dD}}&=&-\frac{5}{18 \pi } e^{-4 \lambda } G \Biggl(e^{6 \lambda } (6 \partial_t\lambda+\eta_c-3 \eta_{\text{N}}+6)+e^{8 \lambda } \eta_{\text{N}}+3 e^{4 \lambda } (-\eta_c+8 \lambda +4)+12 \Bigl(e^{2 \lambda
   } (\partial_t\lambda (6 \lambda -1)\nonumber\\
&+&\lambda  (-\eta_{\text{N}}+12 \lambda -2))-\eta_c \lambda \Bigr) E_1(-6 \lambda )+12 e^{2 \lambda } \Bigl(\partial_t\lambda e^{2 \lambda } (4 \lambda
   -1)+\lambda  \left(-\eta_c+4 \lambda +e^{2 \lambda } (-\eta_{\text{N}}+8 \lambda -2)\right)\Bigr) E_1(-2 \lambda )\nonumber\\
&+&12 \Bigl(\eta_c \lambda +e^{2 \lambda } ((-6 \lambda+1)  \partial_t\lambda+(\eta_c+\eta_{\text{N}}-16 \lambda +2) \lambda )+e^{4 \lambda } ((-4 \lambda +1) \partial_t\lambda+(\eta_{\text{N}}-8 \lambda +2) \lambda )\Bigr) \Gamma (0,-4
   \lambda )\Biggr)\label{etacTT}\\
&+&\frac{1}{72 \pi }e^{-4 \lambda } G \Biggl(12 e^{8 \lambda } \left(24 \lambda ^2-2 (-6 \partial_t \lambda+\eta_{\text{N}}+2) \lambda -2 \partial_t \lambda-\eta_{\text{N}}\right)-e^{6 \lambda } \Bigl(576 \lambda ^2+12 (18
  \partial_t \lambda+\eta_c-3 (\eta_{\text{N}}+2)) \lambda \nonumber\\
&+&30 \partial_t \lambda+7 \eta_c-33 \eta_{\text{N}}+66\Bigr)+6 e^{4 \lambda } (\eta_c+16 \lambda  (3 \lambda -2)+16
 \partial_t \lambda (4 \lambda -1)+\lambda  (-\eta_{\text{N}}+8 \lambda -2)) (\text{Ei}(2 \lambda )-\text{Ei}(4 \lambda ))-4)\nonumber\\
&-&96 \eta_c \lambda  (3 \lambda -1) (\text{Ei}(4 \lambda )-\text{Ei}(6
   \lambda ))+48 e^{2 \lambda } \Bigl(\lambda  (2 (7-6 \lambda ) \lambda +\eta_c(3 \lambda -2)) \text{Ei}(2 \lambda )+\Bigl(2\partial_t \lambda (9 (\lambda -1) \lambda +1)\nonumber\\
&+&\lambda  \left(48 \lambda
   ^2-(3 \eta_c+3 \eta_{\text{N}}+50) \lambda +2 (\eta_c+\eta_{\text{N}}+2)\right)\Bigr) \text{Ei}(4 \lambda )-\Bigl(2 \partial_t \lambda (9 (\lambda -1) \lambda +1)+\lambda  \bigl(36
   \lambda ^2-3 (\eta_{\text{N}}+12) \lambda \nonumber\\
&+&2 (\eta_{\text{N}}+2)\bigr)\Bigr) \text{Ei}(6 \lambda )\Bigr)\Biggr)\label{etacv}\\
&+&\frac{1}{432 \pi }e^{-4 \lambda } G \Biggl(4 e^{8 \lambda } \left(72 \lambda ^2-6 (-6 \partial_t \lambda+\eta_{\text{N}}+2) \lambda -6 \partial_t \lambda-7 \eta_{\text{N}}\right)-e^{6 \lambda } \Bigl(576 \lambda ^2+12 (18
   \partial_t\lambda+\eta_c-3 (\eta_{\text{N}}+2)) \lambda \nonumber\\
&+&126 \partial_t \lambda+23 \eta_c-81 \eta_{\text{N}}+162\Bigr)+18 e^{4 \lambda } (3 \eta_c+4 (4 (\lambda -2) \lambda
   -3)+16 (\partial_t \lambda (4 \lambda -1)\nonumber\\
&+&\lambda  (-\eta_{\text{N}}+8 \lambda -2)) (\text{Ei}(2 \lambda )-\text{Ei}(4 \lambda )))-288 \eta_c (\lambda -1) \lambda  (\text{Ei}(4 \lambda
   )-\text{Ei}(6 \lambda ))\nonumber\\
&-&144 e^{2 \lambda } \Bigl(\lambda  \left(4 \lambda ^2-(\eta_c+10) \lambda +2 \eta_c\right) \text{Ei}(2 \lambda )+\Bigl(-16 \lambda ^3+(-6 \partial_t \lambda+\eta_c+\eta_{\text{N}}+38) \lambda ^2-2 (-7 \partial_t \lambda+\eta_c+\eta_{\text{N}}+2) \lambda \nonumber\\
&-&2 \partial_t \lambda \Bigr) \text{Ei}(4 \lambda )+\Bigl(12 \lambda ^3-(-6
  \partial_t \lambda+\eta_{\text{N}}+28) \lambda ^2+2 (-7 \partial_t \lambda+\eta_{\text{N}}+2) \lambda +2 \partial_t \lambda\Bigr) \text{Ei}(6 \lambda )\Bigr)\Biggr)\label{etacsc}\\
&-&\frac{1}{432 \pi }e^{-4 \lambda } G \Biggl(2 e^{8 \lambda } \left(144 \lambda ^2-12 (-6 \partial_t \lambda+\eta_{\text{N}}+2) \lambda -12 \partial_t \lambda+\eta_{\text{N}}\right)-e^{6 \lambda } (54 \partial_t \lambda (4
   \lambda -1)\nonumber\\
&-&9 (\eta_{\text{N}}-16 \lambda -2) (4 \lambda -1)+\eta_c (12 \lambda -7))-36 e^{4 \lambda } \Bigl(\eta_c-4 \left(2 \lambda ^2+\lambda +1\right)+2 (\partial_t \lambda (4 \lambda
   -1)\nonumber\\
&+&\lambda  (-\eta_{\text{N}}+8 \lambda -2)) (\text{Ei}(2 \lambda )-\text{Ei}(4 \lambda ))\Bigr)-72 \eta_c \lambda  (4 \lambda +1) (\text{Ei}(4 \lambda )-\text{Ei}(6 \lambda ))-72 e^{2 \lambda }
   \Bigl(-\lambda  \left(-8 \lambda ^2+2 \eta_c \lambda +\text{$\eta $c}\right) \text{Ei}(2 \lambda )\nonumber\\
&+&\left(-32 \lambda ^3+2 (-6 \partial_t \lambda+\eta_c+\eta_{\text{N}}-2) \lambda ^2+(-2
 \partial_t \lambda+\eta_c+\eta_{\text{N}}+2) \lambda +\partial_t \lambda\right) \text{Ei}(4 \lambda )\nonumber\\
&+&(\partial_t \lambda (2 \lambda  (6 \lambda +1)-1)+\lambda  (-\eta_{\text{N}} (2 \lambda +1)+4
   \lambda  (6 \lambda +1)-2)) \text{Ei}(6 \lambda )\Bigr)\Biggr).\label{etacphi}
 \label{etac_deDonder}
\end{eqnarray}
\end{widetext}
Numerical values obtained after inserting the corresponding fixed-point values
of the Einstein-Hilbert sector are summarized in Sect.~\ref{sec:results}.

\end{appendix}

\end{document}